\documentclass[9pt,journal]{IEEEtran}
\usepackage{cite}
\usepackage{amsmath,amssymb,amsfonts}
\usepackage{algorithmic}
\usepackage{graphicx}
\usepackage{textcomp}
\usepackage{tikz,xcolor,hyperref}

\usepackage{array}

\usepackage{lipsum}

\newcommand\copyrighttext{%
  \footnotesize \textcopyright 2022 IEEE. Personal use of this material is permitted.
  Permission from IEEE must be obtained for all other uses, in any current or future
  media, including reprinting/republishing this material for advertising or promotional
  purposes, creating new collective works, for resale or redistribution to servers or
  lists, or reuse of any copyrighted component of this work in other works.
  DOI: \href{<https://ieeexplore.ieee.org/document/9667260>}{<DOI No. 10.1109/TAP.2021.3138245>}}
\newcommand\copyrightnotice{%
\begin{tikzpicture}[remember picture,overlay]
\node[anchor=south,yshift=10pt] at (current page.south) {\fbox{\parbox{\dimexpr\textwidth-\fboxsep-\fboxrule\relax}{\copyrighttext}}};
\end{tikzpicture}%
}

\def\BibTeX{{\rm B\kern-.05em{\sc i\kern-.025em b}\kern-.08em
    T\kern-.1667em\lower.7ex\hbox{E}\kern-.125emX}}
\begin{document}
\bstctlcite{IEEEexample:BSTcontrol}

\title{{\fontsize{24}{26}\selectfont{Communication\rule{29.9pc}{0.5pt}}}\break\fontsize{16}{18}\selectfont
Design, Uncertainty Analysis and Measurement of a Silicon-based Platelet THz Corrugated Horn}
\author{Jie Hu, Wei-Tao Lv, Hao-tian Zhu, \IEEEmembership{Member, IEEE},  Zheng Lou, Dong Liu, Jiang-Qiao Ding and Sheng-Cai Shi
\thanks{This Manuscript received on April 11, 2021; revised XXXX, XXXX; accepted XXXX, XXXX. This work was supported in part by the National Key R\&D Program of China (grant No. 2018YFA0404701), the National Natural Science Foundation of China (grant No. 11973095), the Chinese Academy of Science (grants Nos. QYZDJ-SSW-SLH043 and GJJSTD20210002) and NSF of Jiangsu Prov (grant No. BK 20181514). (Corresponding authors: Jie Hu and Sheng-Cai Shi)}
\thanks{J. Hu was with the Key Lab of Radio Astronomy, Purple Mountain Observatory, Chinese Academy of Sciences, Nanjing 210033, China. He is now with Laboratoire Astroparticule and Cosmologie, University Paris Diderot, Paris, 75013, France and also with Observatoire de Paris, University PSL, CNRS, GEPI, Paris 75014, France (e-mail: jiehu@apc.in2p3.fr; jie.hu@obspm.fr ).}
\thanks{W. -T. Lv is with the Key Lab of Radio Astronomy, Purple Mountain Observatory, Chinese Academy of Sciences, Nanjing, 210033 China and also with University of Science and Technology of China, Hefei, 230026, China (e-mail: lvweitao@pmo.ac.cn)}
\thanks{H.-T. Zhu is with the Key Laboratory of Microwave Remote Sensing, Natonal Space Science Center, Chinese Academy of Sciences, Beijing, 100190, China(e-mail:zhuhaotian\_2007@126.com;zhuhaotian@nssc.ac.cn).}
\thanks{J. -Q. Ding is with Jiangsu Key Laboratory of Meteorological Observation and Information Processing, Nanjing University of Information Science and Technology, Nanjing 210044, China and also with Purple Mountain Observatory, Chinese Academy of Sciences and Key Laboratory of Radio Astronomy, Nanjing 210033, China (e-mail: dingjiangqiao@126.com )
 }
 \thanks{Z. Lou, D. Liu and S.-C. Shi are with the Key Lab of Radio Astronomy, Purple Mountain Observatory, Chinese Academy of Sciences, Nanjing 210033, China (e-mail: zhenglou@pmo.ac.cn; dliu@pmo.ac.cn; scshi@pmo.ac.cn ).}
 }

\maketitle
\copyrightnotice
\begin{abstract}
Platelets corrugated horn is a promising technology for their scalability to a large corrugated horn array. In this paper, we present the design, fabrication, measurement and uncertainty analysis of a wideband 170-320 GHz platelet corrugated horn that features with low sidelobe across the band ($<-30$ dB). We also propose an accurate and universal method to analyze the axial misalignment of the platelets for the first time. It is based on the mode matching (MM) method with a closed-form solution to off-axis circular waveguide discontinuities obtained by using Graf addition theorem for the Bessel functions. The uncertainties introduced in the fabrication have been quantitatively analyzed using the Monte Carlo method. The analysis shows the cross-polarization of the corrugated horn degrades significantly with the axial misalignment. It well explains the discrepancy between the designed and the measured cross-polarization of platelets corrugated horn fabricated in THz band. The method can be used to determine the fabrication tolerance needed for other THz corrugated horns and evaluate the impact of the corrugated horn for astronomical observations.
\end{abstract}


\section{Introduction}
\label{sec:introduction}
The terahertz (THz) band, ranging from 0.1 THz to 10 THz, is an important frequency band for radio astronomy and atmospheric science\cite{Siegel2002,Shi2017}. Corrugated horns are frequently used as the feed in THz systems \cite{Gonzalez2012,Shan2012} for their wide bandwidth, low return loss, low sidelobe, low cross-polarization, highly symmetry radiation pattern and high Gaussicity($\approx 98\%$)\cite{Goldsmith1998}. 

Corrugated horns are complicate and traditionally difficult to fabricate into arrays in the THz band\cite{James1984,Sekiguchi2017,Beniguel2005}. For astronomical applications, platelets corrugated horns \cite{Nibarger2011,Zhao2016}, especially silicon-based corrugated horns fabricated by deep reactive iron etching (DRIE) are favored for their high fabrication accuracy, scalability into a large corrugated horn array and their thermal compatibility with silicon-based detectors like transition edge sensor (TES)\cite{Thornton2016} 
and microwave kinetic inductance detectors (MKIDs) \cite{Day2003,Hu2020} at low temperature. Such corrugated horn array is of particular interest in the applications such as the detection of B-mode \cite{Thornton2016}, where a large number of low cross-polarization antennas are required. 

Corrugated horns are delicate and are quite sensitive to fabrication uncertainties \cite{Gonzalez2012}. There are several uncertainties related to the fabrication of the silicon-based platelet corrugated horn. The first two uncertainties are the underetching effect and the non-uniformity of etching depth of each platelet \cite{Nibarger2011}, both of which are related to the DRIE process. The third is the axial misalignment between the platelets during the assembly. It is not considered for corrugated horns fabricated by direct machining \cite{Sekiguchi2017} or electroforming \cite{Beniguel2005}. For the state-of-art fabrication, it is considered to be small as the horn is axial-symmetrically milled or eletroformed \cite{Gonzalez2012}. However, such uncertainty has to be taken into consideration for THz platelet corrugated horns, as the wavelength $\lambda$ in THz band is very short, making it comparable with the wavelength. It will not only introduce extra cross-polarization as the discontinuity between non-axially connected circular waveguides gives rise to all degenerate modes in the horn but also reduces the Gaussicity. This is one of the possible reason that the cross-polarization of the state-of-art platelets corrugated horns is limited to be on the order of $-20$ dB at THz band \cite{Nibarger2011,Thornton2016}.
\begin{figure}
    \centering
    \includegraphics[width=\columnwidth]{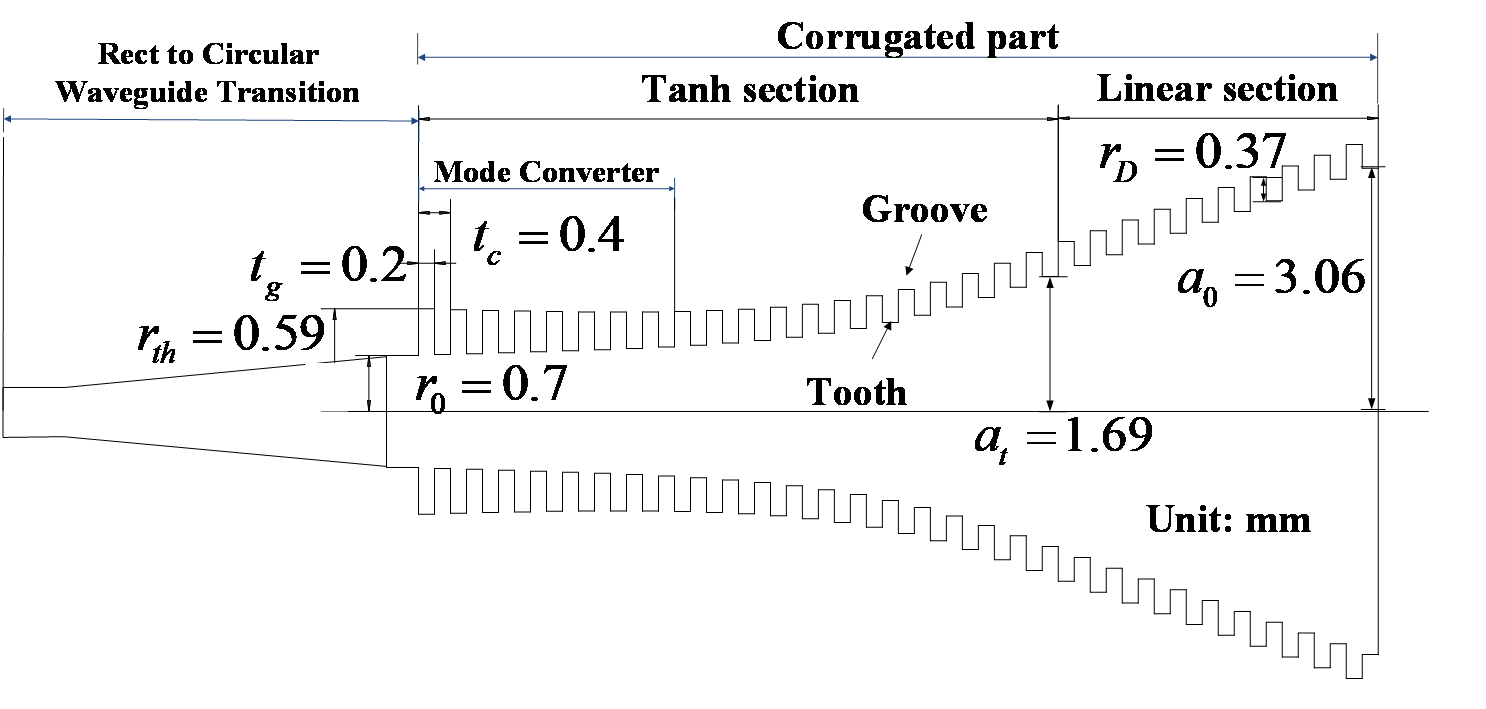}
    \includegraphics[width = \columnwidth]{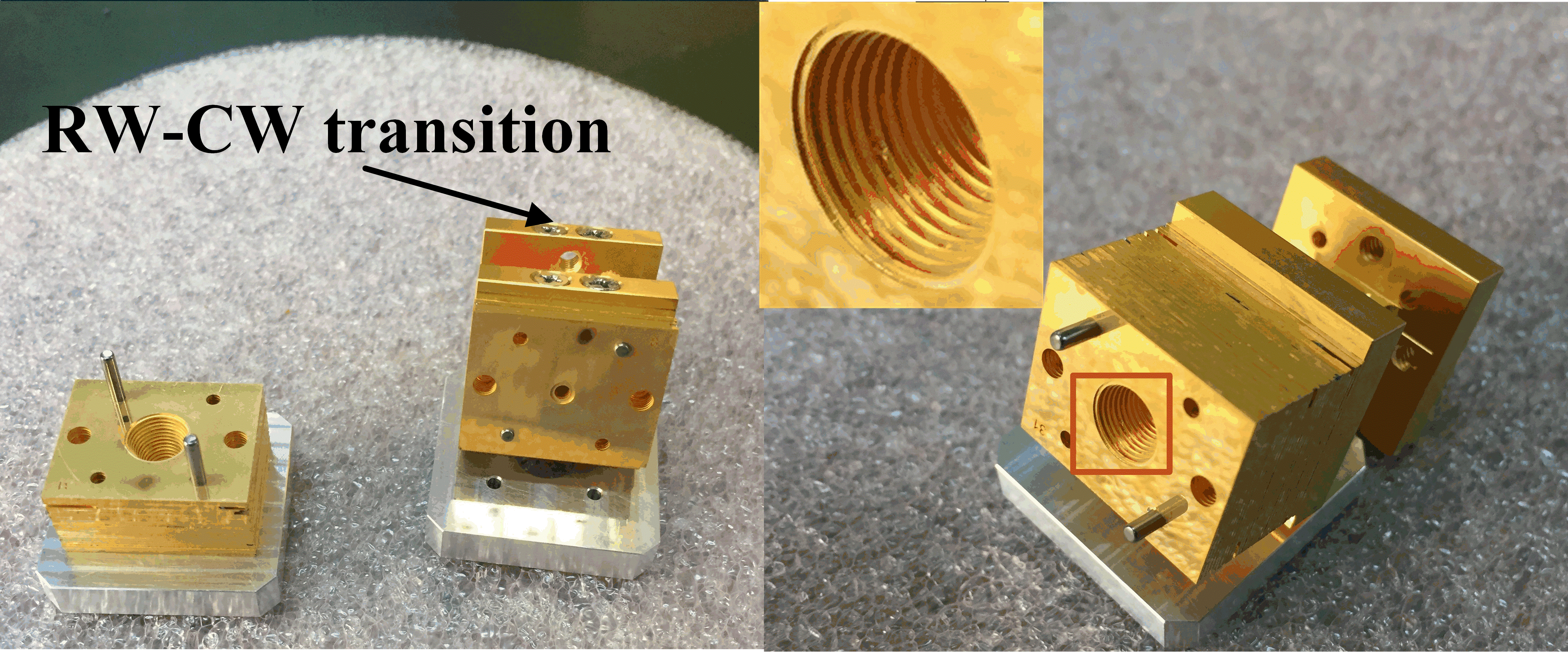}
    \caption{Designed and fabricated horn. Upper: Dimension of the designed tanh + linear profiled corrugated horn. It is segmented into four parts, the rectangular to circular waveguide (RW-CW) transition, the mode converter, the tanh-profiled section and the linear profiled section. Lower: Fabricated silicon platelets and the assembled corrugated horn.The size of each platelet is 16 x 20 $mm^2$. }
    \label{fig:horn_dimension}
\end{figure}

The uncertainty introduced by the misalignment between the platelets is rather difficult to model in commercially available software based on the finite element method (FEM) or the method of momentum (MoM). It takes a huge amount of meshes to analyze the horn to such precision ($\lambda$/100) and it takes months to fully characterize the effect of the uncertainty \cite{Gonzalez2012}. For the MM method based software like CORRUG \cite{McKay2013}, only axially symmetrical horns can be analyzed. To authors’ knowledge, until now, no uncertainty analysis has been performed the corrugated horns.

Corrugated horns have been analyzed by the MM method for decades \cite{James1984}. In this work, we analyze asymmetrical corrugated horns with certain misalignment between corrugations for the first time using the MM method. It is based on a closed-form analytical solution to asymmetrical circular waveguide discontinuities obtained by using Graf addition theorem for Bessel function \cite{Shen1995}. It can analyze the uncertainty introduced by the misalignment between the platelets efficiently to decent accuracy as long as sufficient modes are considered. This paper is organized as follows. In Section II, the design, fabrication and measurements of the horn is introduced. In Section III, we present the analytical solution to the off-axis circular waveguide discontinuity as well as the uncertainties introduced by the misalignment between the platelets and by the non-uniformity of the etching depth obtained by the Monte Carlo analysis\cite{Gonzalez2012}. We have also compared the measured results with the other state-of-art corrugated horns.

\section{Horn Design, Fabrication and Characterization}\label{sec:design}
\subsection{Corrugated Horn Design}
Profiled corrugated horns are preferred as they are more compact and have lower sidelobe than the conventional conical ones. The tanh + linear profile is chosen for low sidelobe \cite{McKay2013}. The profile of tanh part of the corrugated horn is expressed as 
\begin{align}
r(z) = r_0 + (a_0 - r_0){\frac{(1-A)z}{L_{profile}}} + \frac{A}{2}[\mathrm{tanh(\frac{B\pi z}{2L_{profile}+ 1})}],
\end{align}
where $r_0$ is the radius of the input circular waveguide, $a_0$ is the radius of the circular waveguide at the end of the tanh profile, $L_{profile}$ is the length of the tanh profile, A and B are two parameters for optimization. 

Instead of pursuing for ultimate high Gaussicity, we optimize the horn to be wideband, low cross-polarization and low sidelobe. The depth of the grooves in the mode converter is exponentially tapered from around half wavelength to quarter wavelength \cite{Shan2012}, which is expressed as $r_D(z)=r_{th}e^{-bn}\text{,}$
where $r_{th}$ is the depth of the first groove and n is the number of the corrugation in the mode converter. n is set to be 8 in our simulation. The optimized dimension of the corrugated horn is shown in Fig. \ref{fig:horn_dimension}.

The thickness of each corrugation $t_c$ is set to be 400 $\mu m$ and the depth of each groove $t_g$ is set to be 200 $\mu m$ to facilitate the fabrication of all the corrugations on a single silicon wafer. The number of corrugations in the tanh profile section and the linear section is set to be 20 and 10 respectively. The length of the rectangular waveguide to circular waveguide (RW-CW) transition is set to be 5 mm. The input rectangular waveguide is chosen to be WR-4.3 to reduce the return loss at low frequency. 

\begin{figure}
    \centering
    \includegraphics[width = \columnwidth]{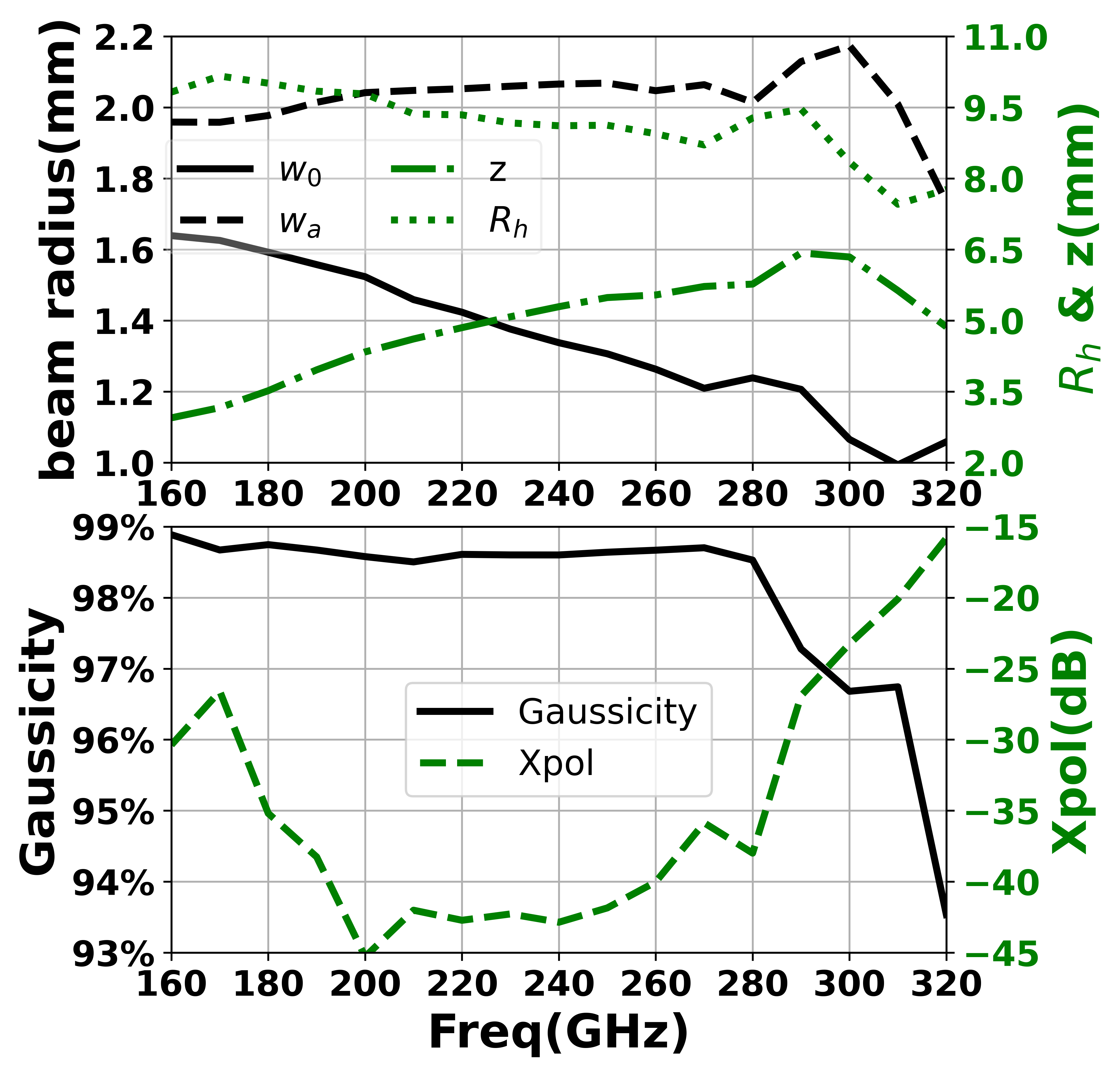}
    \caption{The simulated main parameters for the corrugated horn. Upper: the beam waist ($w_0$), the beam radius ($w_a$) on the aperture, the distance between the beam waist to the aperture and the radius of curvature on the aperture (Rh); Down: The Gaussicity and cross-polarization, The Gaussicity is obtained by fitting the electrical field at the aperture to the fundamental gaussian mode.} 
    \label{fig:horn parameters}
\end{figure}

The simulated Gaussian beam parameters is shown in Fig. \ref{fig:horn parameters}. It can be seen that the beam radius $w_a$ and the radius of curvature $R_h$ on the horn aperture are almost constant from 160 - 280 GHz. The beam radius is around 2 mm on the horn aperture, about $0.66a_0$, which is similar to linear conical corrugated horns \cite{Goldsmith1998}. The simulated cross-polarization is less than 20 dB in the band of 160-310 GHz. To achieve Gaussicity better than 97 \%, the cross-polarization should be controlled to be below -25 dB.

\subsection{Fabrication and Assembly}
Only the corrugation part of the corrugated horn is fabricated by silicon platelets. The horn was fabricated on a 6-inch silicon wafer. There were two DRIE steps in our process. The first etched 400$\ \mu m$ silicon to form the center holes and the alignment pin holes and the second etched 200 $\mu m$ silicon to form the corrugations. A layer of 1.2 $\mu m $ gold was deposited on the surface of the silicon to minimize the loss. The depths of the corrugation range from 185 $\mu m$ to 215 $\mu m$ and the sidewall angles to the horizontal is around $88.5^o$ due to the underetching effect.

The fabricated silicon platelets are shown in Fig. \ref{fig:horn_dimension}.  All the platelets are aligned by two stainless alignment pins. The alignment accuracy is estimated to be around 10-20 $\mu m$, limited by the accuracy of the alignment pin and the underetching effect. The platelets were assembled and pressed tightly between two aluminum plates to minimize the air gap between each adjacent platelet. Then, these platelets were glued with conductive epoxy around the four sides. 

%

\subsection{Measurements}

\begin{figure}
    \centering
    \includegraphics[width = \columnwidth]{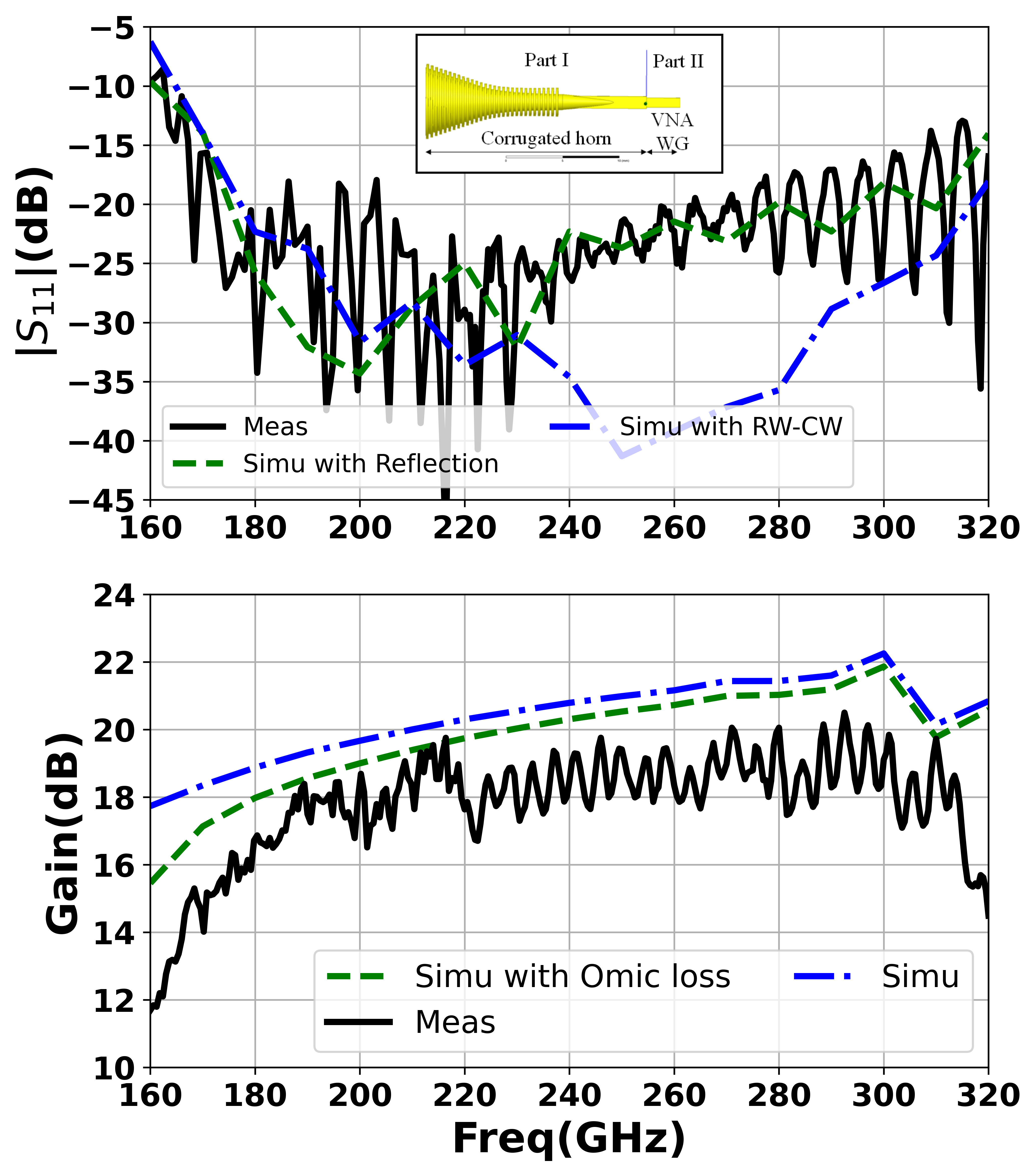}
    \caption{Measured return loss and Gain comparison with simulation. Upper: measured and simulated return loss; Lower: measured and simulated gain. All the simulations are done in HFSS to include the RW-CW transition. The simulation is done within 160-220 GHz with WR-5.1 waveguide and in 220-320 GHz with WR-3.4 waveguide introduced by the VNA}
    \label{fig:measured gain and s11}
\end{figure}

The return loss and the gain of the corrugated horn are measured by the THz vector network analyzer (THz VNA) at the State Key Laboratory of Millimeter Waves, City University of Hong Kong \cite{Zhu2016} and are compared with simulation, as is shown in Fig. \ref{fig:measured gain and s11}. The THz VNA consists of a Keysight N5245A PNA-X with two OML extension modules. The horn is measured in two bands, namely 160-220 GHz (WR-5.1) and 220-320 GHz (WR-3.4) defined by the OML extension modules. Microwave absorption materials are placed around the corrugated horn aperture to minimize the multi-reflection in free space. It can be seen that the measured $S_{11}$ is higher than the simulated one especially in the band of 220-320 GHz because of the mis-match between input waveguide of the corrugated horn (WR-4.3) and that of THz VNA (WR-3.4). When taking it into consideration, the measured S11 shows great agreement with the simulated one. The measured gain is around 2 dB less than the simulated result with perfect electric conductor on average. The loss mainly comes from the omic loss in the horn and the loss from the long rectangular waveguide in the RW-CW transition, which is estimated to be approximately 0.6 dB. 

\begin{figure}
    \centering
    \includegraphics[width = \columnwidth]{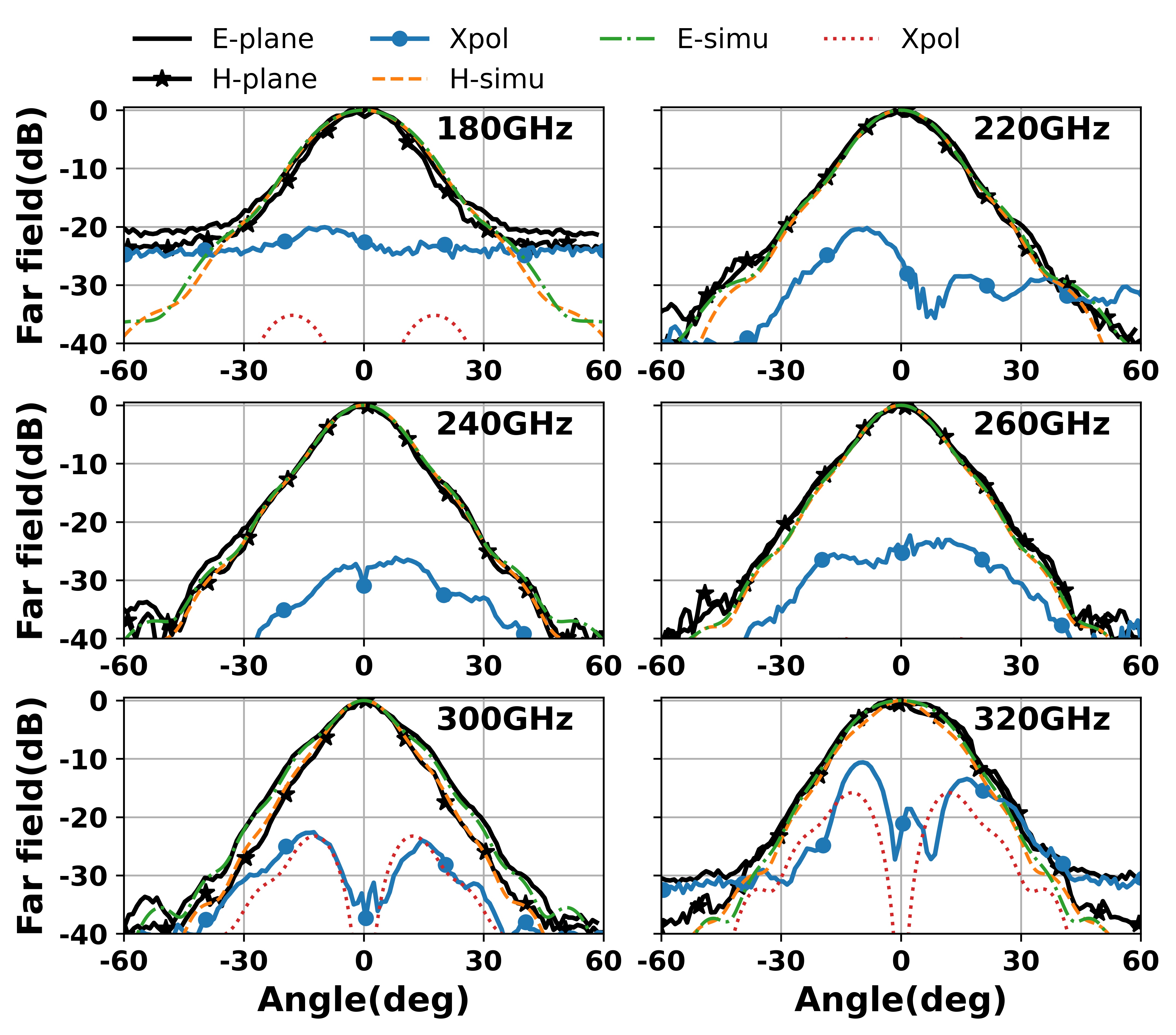}
    \caption{ Measured far-fields comparison with simulation. The dynamic range of the system is limited by the SIS mixer below 220 GHz, which is around 20 dB at 180 GHz.}
    \label{fig:measured farfield}
\end{figure}

The simulated and measured far-field of the horn is shown in Fig. \ref{fig:measured farfield}.  It is measured in the beam measurement system based on a superconducting-insulator-superconductor (SIS) mixer \cite{Lou2014,Li2019,Hu2017} at Purple Mountain Observatory (PMO). The beam is highly symmetric in the band of 180-320 GHz and the sidelobe is lower than -30 dB across the band. The cross-polarization is measured in the $45^o$ D-plane,limited by the system. The measured cross-polarization is below -20 dB in the band of 180-310 GHz, but it is much higher than the simulated results.  

\section{Uncertainty Analysis} \label{sec:uncertainty}
Uncertainty analysis based on closed-form MM method is performed to investigate the origin of the measured high cross-polarization. There are four uncertainties introduced during the fabrication of the silicon based corrugated horn, misalignment between platelets and the gap between the platelets, the underetching effect, non-uniform etching depth,  which are shown in Fig. \ref{fig:offset waveguide}. The underetching effect and the non-uniform etching depth are related to the process of DRIE. The misalignment between plateles and the gap between the platelets are introduced during the assembly of the corrugated horn. The uncertainty introduced by the underetching effect is negligible and that by the gap between the platelets is not considered in this work. 

\subsection{Solution to Off-axis Circular Waveguide Discontinuity}\label{subsec:theory}

\begin{figure}
    \centering
    \includegraphics[width = \columnwidth]{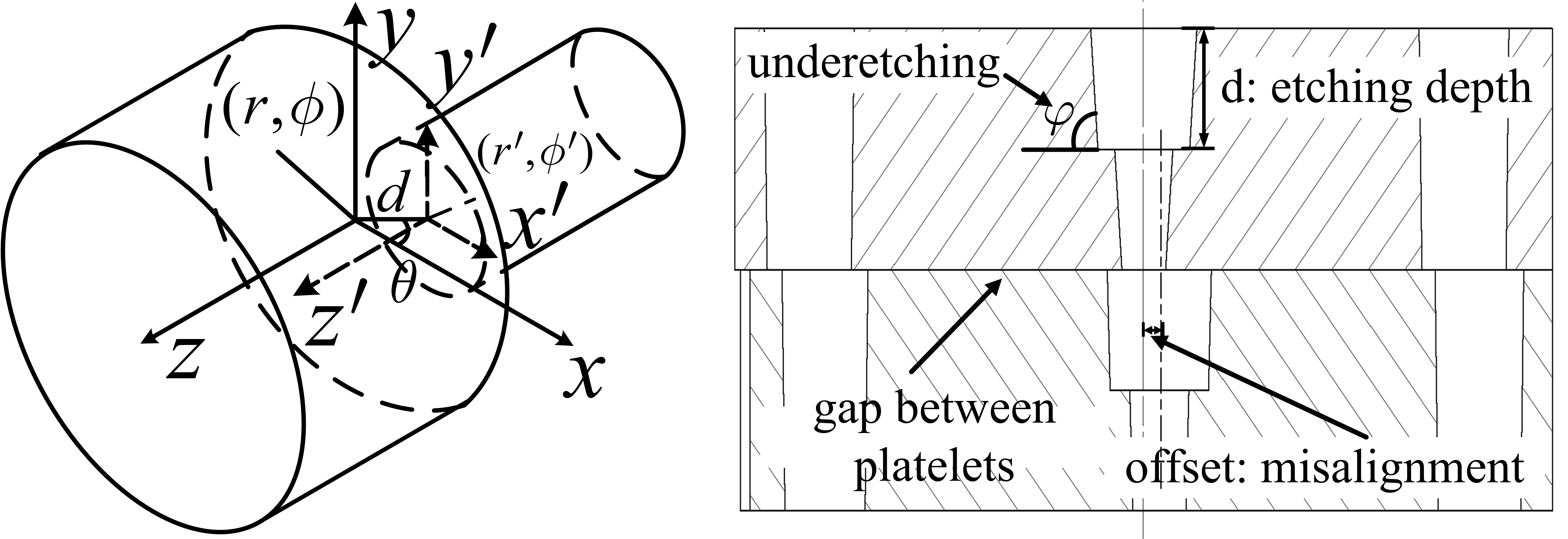}
    \caption{Left: Illustration of the uncertainties introduced by the fabrication of the silicon based corrugated horn. Right: Illustration of the off-axis circular waveguide connection. }
    \label{fig:offset waveguide}
\end{figure}

As is shown in Fig. \ref{fig:offset waveguide}, in a cylinder coordinate ($r,\phi$), the longitudinal components for the circular waveguide are expressed as
\begin{align} \label{eq:mode}
    \begin{aligned}
    \varphi_{mn}^{TE} &= N_{mn}^{TE}J_m(k_{mn}^{TE}r) \left(
        \begin{aligned}
            & \mathrm{sin}(\theta) \\
            & \mathrm{cos}(\theta) 
        \end{aligned}\right)
        \\
    \varphi_{mn}^{TM} &= N_{mn}^{TM}J_m(k_{mn}^{TM}r)\left(
    \begin{aligned}
            & \mathrm{sin}(\theta) \\
            & \mathrm{cos}(\theta) 
        \end{aligned}\right) \\
    \end{aligned},
\end{align}
where $N_{mn}^{TE}$ and $N_{mn}^{TM}$ are the normalization constant, $J_m$ is the $m^\text{th}$ order Bessel function of the first kind, $k_{mn}^{TE}=x^\prime_{mn}/R$ and $k_{mn}^{TM} = x_{mn}/R$ are the cutoff wavenumber of the mode, $R$ is the radius of the circular waveguide, $x_{mn}$ and $x_{mn}^\prime$ are the $n^\text{th}$ root of $J_m(x)=0$ and $J_m^\prime(x)=0$ respectively. 

For a circular waveguide connection with certain offset from the axis, by Graf addition theorem for Bessel function \cite{Shen1995}, the longitudinal components can be expressed as 
\begin{small}
\begin{align}\label{eq:offsetsolution}
    J_m(\alpha r) =\left(
    \begin{aligned}
            & \mathrm{sin}(\theta) \\
            & \mathrm{cos}(\theta) 
        \end{aligned} \right)
     \sum_{k=-\infty}^{\infty}J_{k-m}(\alpha d) J_k(\alpha r^\prime)\left(
    \begin{aligned}
            & \mathrm{sin}(k\phi^\prime - (k-m)\theta) \\
            & \mathrm{cos}(k\phi^\prime - (k-m)\theta) 
        \end{aligned} \right),
\end{align}
\end{small}
where $\alpha$ is an arbitrary constant, $(d,\theta)$ are the coordinate of the center of the smaller waveguide. The series in equation \eqref{eq:offsetsolution} converges quickly and terms with large k can be neglected in simulation. It is valid only when the small waveguide should be within the large waveguide. The mode coupling of the TE and TM modes between the large and the small waveguide can be expressed as
\begin{align}\label{eq:Modecoupling}
    \begin{aligned}
    M_{hh}[m,n,m^\prime,n^\prime] & = (k_{m^\prime n^\prime}^{TE})^2 \iint_S \varphi_{mn}^{TE}\varphi_{m^\prime n^\prime}^{TE}ds\\
    M_{ee}[m,n,m^\prime,n^\prime] & = (k_{m^\prime n^\prime}^{TE})^2 \iint_S \varphi_{mn}^{TE}\varphi_{m^\prime n^\prime}^{TE}ds\\
    M_{eh}[m,n,m^\prime,n^\prime] & = -\oint_c \varphi_{mn}^{TE}\mathbf{e}_{m^\prime n^\prime}^{TE}\cdot d\hat{l}\\
    M_{he}[m,n,m^\prime,n^\prime] & = 0
    \end{aligned},
\end{align}
where $M_{hh}$, $M_{ee}$, $M_{eh}$ and $M_{he}$ are the coupling of the TE-TE, TM-TM, TM-TE and TE-TM modes between the two waveguides. $\mathbf{e}_{m^\prime n^\prime}^{TE} = \hat{z}\times \nabla \varphi_{m^\prime n^\prime}^{TE},$,The third integal is done on the contour of the small waveguide. 

By substituting equation \eqref{eq:mode} and \eqref{eq:offsetsolution} into equation \eqref{eq:Modecoupling}, the off-axis circular waveguide discontinuity can be solved analytically. Detailed closed-form solution can be found in \cite{Shen1995}.

\subsection{The Misalignment between Platelets}
\begin{figure}
    \centering
    \includegraphics[width = \columnwidth]{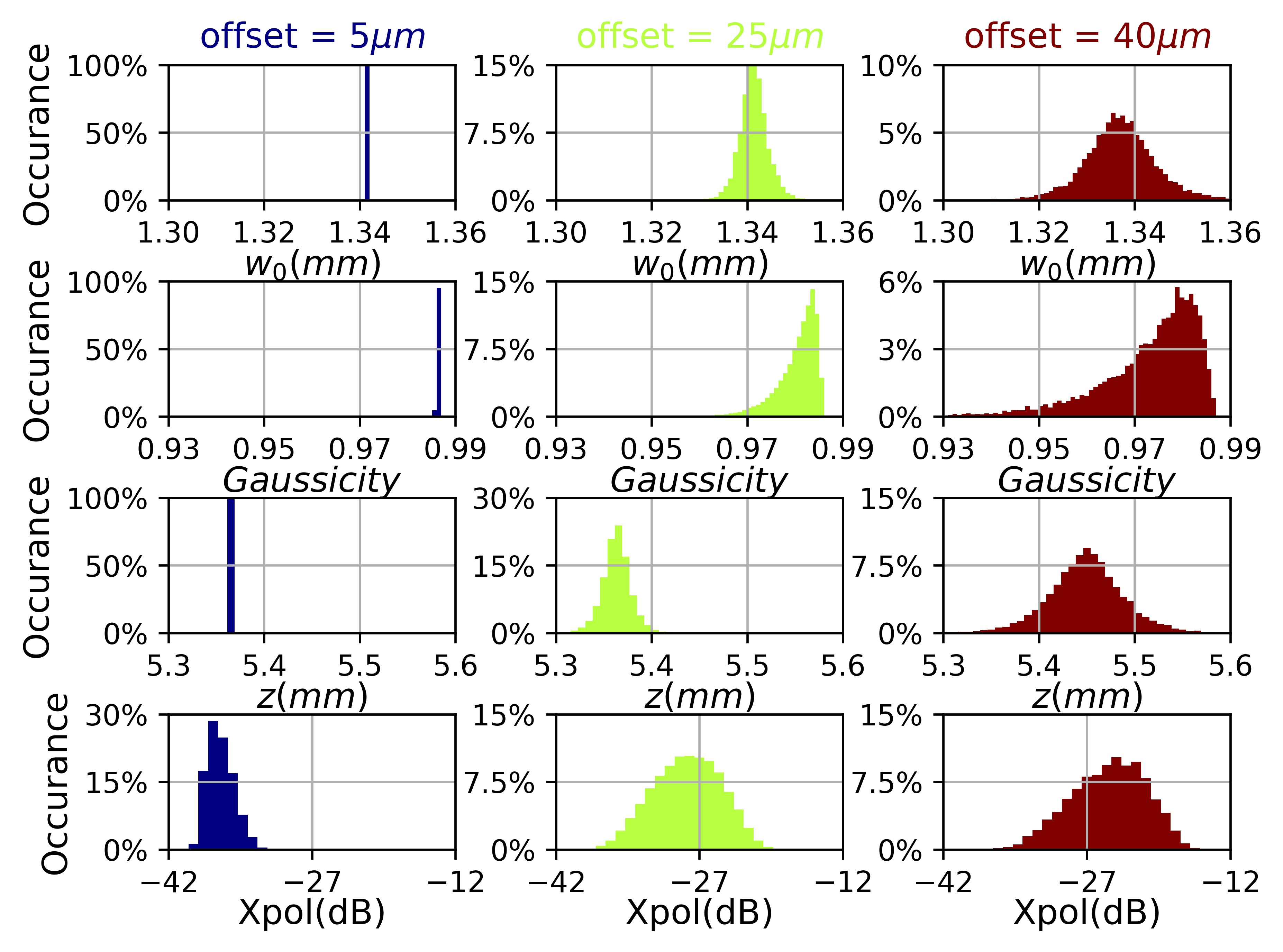}
    \caption{Statistics for different misalignment uncertainty between the platelets for the corrugated horn at 240 GHz. In total, 10,000 simulations has been performed.}
    \label{fig:mis-alignment error at 240GHz}
\end{figure}

The uncertainty from misalignment between the platelets originates from the uncertainty in the size of the alignment pin hole and that in the diameter of the alignment pin itself. We assume the uncertainty in the axial distance of each platelet to the axis of the corrugated horn is uniformly distributed. Tolerances with maximum deviation of $5-40 \ \mu m$ have been analyzed, as is shown in Fig. \ref{fig:mis-alignment error at 240GHz}. In total, 90 $TE_{\text{mn}}$ and 90 $TM_{\text{mn}}$ modes with $m<8$ and $n<20$ into consideration, as the terms with $m \le 8$ is negligible as the series in equation \eqref{eq:offsetsolution} converges quickly. 
As the uncertainty increases, the uncertainty of the main parameters of the horn increases. The cross-polarization is most sensitive to the misalignment of the platelets. There is around 10 dB degradation if the misalignment between the platelets is $25 \ \mu m$ ($\lambda$/50 @240GHz). The Gaussicity also degrades significantly as it is closely related to the cross-polarization. It should be noted that there is no case with lower cross-polarization than the designed value (-42dB) in simulations with misalignment between platelets. This is because the symmetry of the horn is destroyed and almost the mode that can propagate in the horn will be excited including the degenerated modes of $TE_{\text{1n}}$ and $TM_{\text{1n}}$, according to equation \eqref{eq:offsetsolution}. Thus, the energy in the cross-polarization increases and the Gaussicity decreases.

\begin{figure}
    \centering
    \includegraphics[width = \columnwidth]{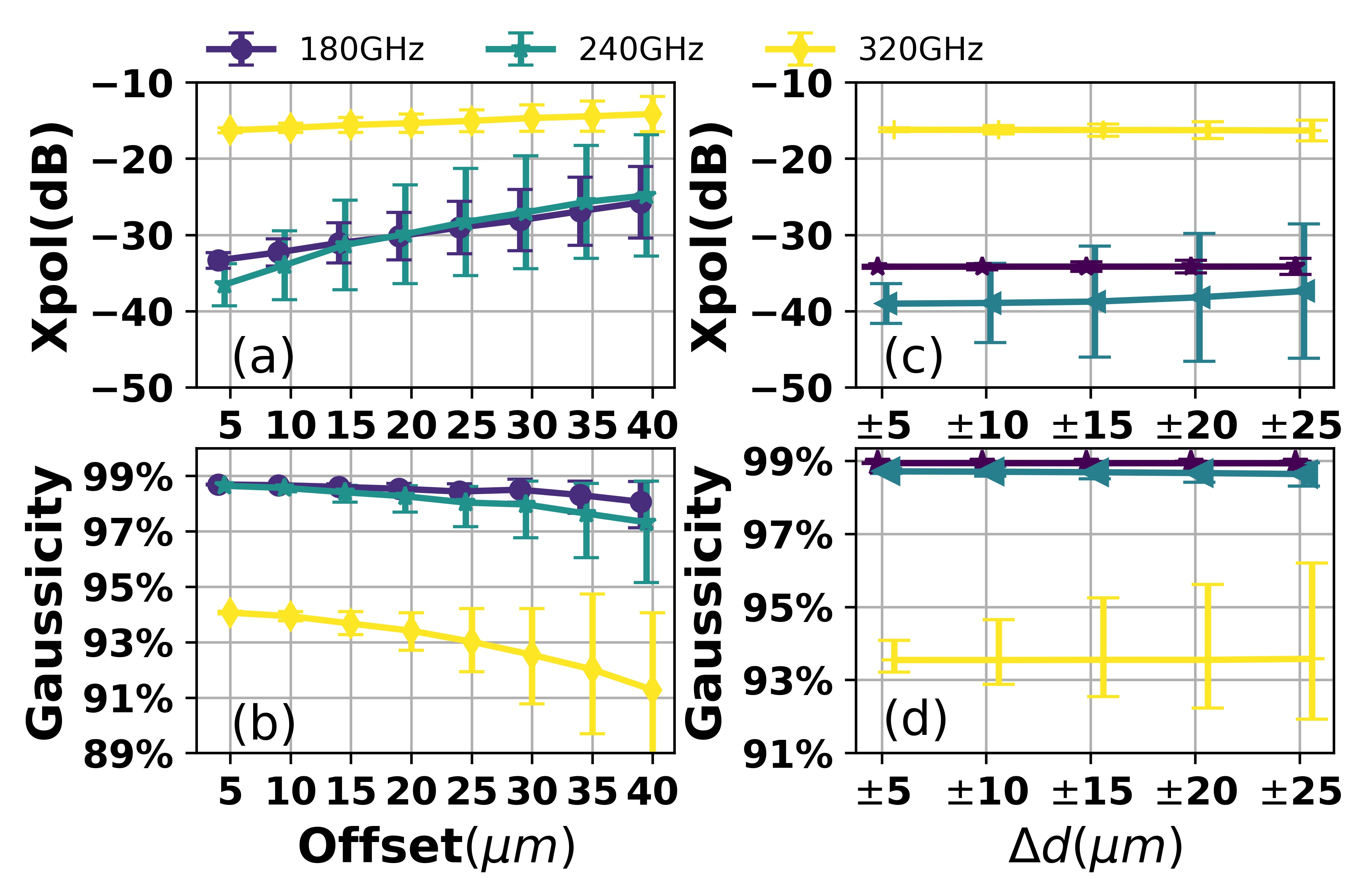}
    \caption{Uncertainty in cross-polarization and Gaussicity  from misalignment and etching depth for different frequencies. (a) Cross-polarization uncertainty from misalignment. (b) Gaussicity uncertainty from misalignment. (c) Cross-polarization uncertainty from etching depth. (d) Gaussicity uncertainty from etching depth. The statistics for all the parameters are fitted as normal distribution. The curves are plot with certain horizontal offset to make it clear. The upper boundary for the Gaussicity is given as the maximum its statistics doesn’t observe norm distribution. 10,000 independent simulations are performed for each tolerance.}
    \label{fig:misalignment at for different freq}
\end{figure}

\begin{table*} 
\caption{Comparison of Corrugated Horns}\label{table:compare}

\begin{center}
\renewcommand\arraystretch{1.5}
\begin{tabular}{c c c c c c c c}

 \hline
             & Frequency(GHz) & Bandwidth &Sidelobe & Cross-pol & Return loss & Corrugation Type & Method\\ 
 \hline\hline
Sekiguchi et al\cite{Sekiguchi2017}   & 120-270 GHz    & 2.25      & $<-25$ dB   & $<-20$ dB    & $<-15$ dB      & Linear           &Direct machined \\
 \hline
Béniguel et al\cite{Beniguel2005}    & 75-110 GHz      & 1.47     & $<-50$ dB   & $<-40$ dB    & $<-20$ dB      & Linear           & Electroforming  \\ 
 \hline
McKay et al \cite{McKay2013}       & 84-104 GHz    & 1.23      & $<-35$ dB     & $<-35$dB    & $<-30$ dB      & Tanh             & Electroforming \\ 
 \hline
Datta et al \cite{Thornton2016}      & 70-175 GHz    & 2.33      & $<-25$ dB   & $<-20$ dB    & N/A            & Ring-loaded Linear     & MEMS + Au coated \\ 
\hline
Nibarger et al \cite{Nibarger2011}      & 120-170 GHz    & 1.41      & $<-30$ dB   & $<-20$ dB    & $<-20$ dB            & sin     & MEMS + Au/Cu coated \\ 
\hline
This work         & 170-320 GHz    & 1.88      & $<-30$ dB   & $<-20$ dB    & $<-15$ dB        & Tanh            & MEMS + Au coated \\  

 \hline
\end{tabular}
\end{center}
\end{table*}

\begin{figure}
    \centering
    \includegraphics[width = \columnwidth]{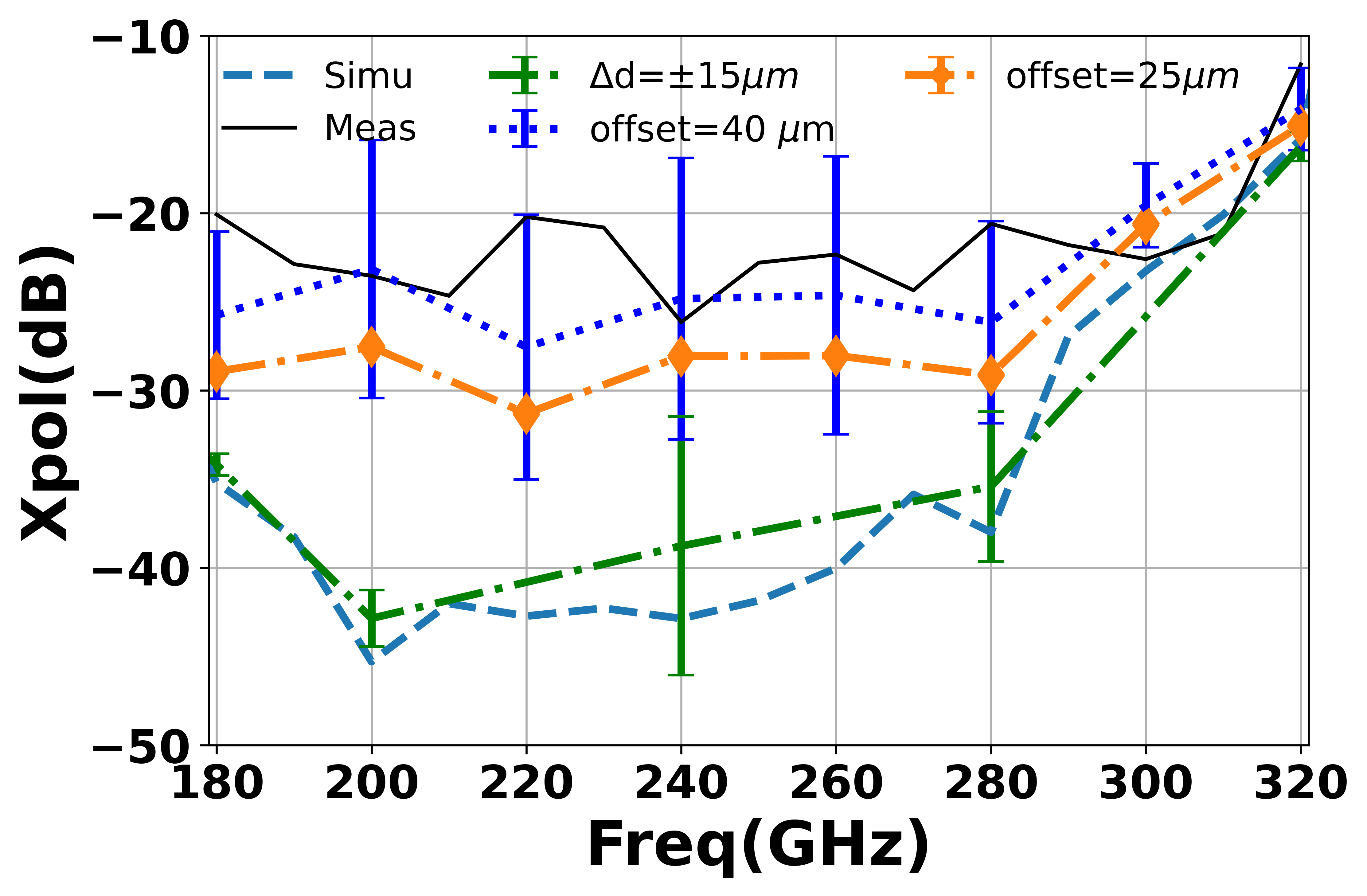}
    \caption{Cross-polarization comparison between measurements and simulation in the band of 180-320 GHz. The error bar is given as 2$\sigma$. The uncertainty in the etching depth is $\pm15 \ \mu m$. The cross-polarization increases quickly as the uncertainty in the misalignment between platelets increases.}
    \label{fig:cross-polarization}
\end{figure}

The statistics the uncertainty introduced by the misalignment at different frequencies is compared in Fig. \ref{fig:misalignment at for different freq}(a) and (b). All the statistics are fitted to be normal distribution. The error bar is set at $2\sigma$, $\sigma$ is the standard deviation. The upper boundary for the Gaussicity is set to be the maximum value in the simulation, as the Gaussicity does not observe the normal distribution. It can be seen that the misalignment purely degrades the performance of the horn.especially the cross-polarization and the Gaussicity. With the uncertainty in the misalignment increasing, the average performance of the horn degrades and the standard deviation increases.

\subsection{Non-Uniformity of the Etching Depth}
The main uncertainty introduced by the DRIE process is the non-uniformity of the etching depth, which is around $\pm 15\ \mu m$ in our case. Tolerances of $\pm5\ \mu m-\pm 25\ \mu m$ are simulated as is shown in Fig. \ref{fig:misalignment at for different freq} (c) and (d). The uncertainty in the etching depth is assumed to be uniformly distributed as well. There is a certain amount of simulation that has lower cross-polarization and better Gaussicity than the designed value. It means the horn can be further optimized by varying the etching depth in each corrugation. It can be seen that as the uncertainty increases, the mean value remains the same but the standard deviation of the parameters increases. This is quite different compared with the case in the uncertainty introduced by the mis-alignment, where the uncertainty increases, the mean performance of the horn degrades.


\subsection{Cross-polarization analysis and dicussion}

The measured cross-polarization is compared with the simulated results in Fig. \ref{fig:cross-polarization}. The analysis shows the cross-polarization is mainly dominated by the axial misalignment between the platelets, while the uncertainty introduced by the non-uniformity of the etching depth is negligible. The misalignment between the platelets is estimated to be around 25-40 $\mu m$, which is much larger than our initial estimation. It is mainly limited by the accuracy of the diameter of the holes for the alignment pins, as the hole is fabricated in just one DRIE process.

The performance of the designed horn is compared with other publications in Table \ref{table:compare}. The state-of-art of the traditionally fabricated corrugated horn of shows better cross-polarization. The corrugations in the horn fabricated by direct machining or electroforming are intrinsically aligned as they are milled along the same axis. The cross-polarization of the corrugated horn formed by platelets is limited to be around -20 dB in the THz band due to the axial alignment The horn presented in this work features with wide bandwidth without ring-loaded structure\cite{Takeda1976} and a low sidelobe over the whole bandwidth.

\section{Conclusion}
A wideband low sidelobe profiled corrugated horn based on silicon platelets is designed, fabricated and characterized. The measured beam pattern is in agreement with simulation and the measured sidelobe is lower than -30 dB across the band. The measured gain of the corrugated horn is approximately 18 dB and the measured return loss is basically less than -15 dB across the band of 170-320 GHz. 
The misalignment between the platelets degrades the cross-polarization and Gaussicity. It is estimated between $25 \ \mu m - 40 \ \mu m$ and dominates the cross-polarization of the horn. The uncertainty of the axial misalignment between the platelets should be less than $\lambda/100$ to achieve a cross-polarization level of -30 dB for this horn.

\section*{Acknowledgment}
The authors would like to thank J. P. Yang, Z. H. Lin and Q. J. Yao of PMO, Y. Liu of Beijing Key Laboratory of Millimeter and Terahertz Technology, Beijing Institute of Technology for helpful discussion. The return loss and the gain of the horn were measured in the State Key Laboratory of Millimeter Waves, City University of Hong Kong, Hong Kong. 


\bibliographystyle{IEEEtran}

\bibliography{reference}
\end{document}